\documentclass[usenatbib,galley,onecolumn]{mnras}

\usepackage{graphicx}	% Including figure files
\usepackage{amsmath}	% Advanced maths commands
\usepackage{amssymb}	% Extra maths symbols
\usepackage{multicol}        % Multi-column entries in tables
\usepackage{bm}		% Bold maths symbols, including upright Greek
\usepackage{pdflscape}	% Landscape pages

\usepackage[T1]{fontenc}
\usepackage{ae,aecompl}

\title[Water ice erosion by low energy ions]{{A laboratory study of water ice erosion by low energy ions }}

\author[E. A. Muntean et al.]{Elena A. Muntean$^{1}$\thanks{Contact e-mail: \href{mailto:emuntean01@qub.ac.uk}{emuntean01@qub.ac.uk}},
Pedro Lacerda$^{1, 2}$
Thomas A. Field$^{3}$, Alan Fitzsimmons$^{1}$,
\newauthor Wes C. Fraser$^{1}$, Adam C. Hunniford $^{1}$ and Robert W. McCullough $^{1,3}$\\
$^{1}$ Astrophysics Research Centre, School of Mathematics and Physics, Queen`s University Belfast, BT7 1NN, UK \\
$^{2}$ Max Planck Institute for Solar System Research, G\"{o}ttingen, 37191, Germany\\
$^{3}$ Centre for Plasma Physics, School of Mathematics and Physics, Queen`s University Belfast, BT7 1NN, UK }

\date{Accepted for publication at MNRAS, 2016}

\pubyear{2016}

\begin{document}
\label{firstpage}
\pagerange{\pageref{firstpage}--\pageref{lastpage}}
\maketitle

\begin{abstract}
Water ice covers the surface of various objects in the outer solar system. Within the heliopause, surface ice is constantly bombarded and sputtered by energetic particles from the solar wind and magnetospheres. We report a laboratory investigation of the sputtering yield of water ice when irradiated at 10 K by 4 keV singly ($^{13}$C$^+$, N$^+$, O$^+$, Ar$^+$) and doubly charged ions ($^{13}$C$^{2+}$, N$^{2+}$, O$^{2+}$). The experimental values for the sputtering yields are in good agreement with the prediction of a theoretical model. There is no significant difference in the yield for singly and doubly charged ions. Using these yields, we estimate the rate of water ice erosion in the outer solar system objects due to solar wind sputtering. Temperature programmed desorption of the ice after irradiation with $^{13}$C$^+$ and $^{13}$C$^{2+}$ demonstrated the formation of $^{13}$CO and $^{13}$CO$_2$, with $^{13}$CO being the dominant formed species.
\end{abstract}


%%%%%%%%%%%%%%%%%%%%%%%%%%%%%%%%%%%%%%%%%%%%%%%%%%

%%%%%%%%%%%%%%%%% BODY OF PAPER %%%%%%%%%%%%%%%%%%

% The MNRAS class isn't designed to include a table of contents, but for this document one is useful.
% I therefore have to do some kludging to make it work without masses of blank space.
%\begingroup
%\let\clearpage\relax
%\tableofcontents
%\endgroup
%\newpage

\section{Introduction}

Water ice is a major component of solid bodies in the outer solar system and played a major role in the planet formation process  \citep{Lunine}. Beyond $\sim4$ AU it is not warm enough for water ice to undergo substantial sublimation, resulting in potential lifetimes of order of the age of the solar system and its observed presence on the surfaces of moons and Kuiper Belt objects e.g. \cite{Guilbert2009}, \cite{Dalton}, \cite{Brown2012}. Water ice, however, will still undergo erosion due to  dust and macroscopic body impact and irradiation by cosmic rays, ions in the solar wind and, for moons and ring particles, magnetospheric ions. Chemical species can be lost from or redistributed on surfaces as a result of this bombardment and new chemical species can be formed \citep{Hudson, Johnson1990}. Some of the newly formed species  will remain embedded in the ice, changing its chemical composition, whilst others may in turn be ejected from the surface. This process has been observed in laboratory ion irradiation experiments involving ices e.g. \cite{Brunetto2006}. In some cases sputtering leads to the formation of thin atmospheres made of water molecules, its dissociation products plus the sputtering species; these have been observed around Jovian and Saturnian Satellites e.g. \citep{Hall95,Brown1996,Johnson2008}. Additionally, energetic ion sputtering is at least partly responsible for global compositional changes across Tethys and Mimas \citep{Schenk2011}. 

Sputtering of H$_2$O ice has been intensively studied in laboratory experiments with both low energy ions (below 10 keV) and high energy ions (from 10 keV to a few MeV). Almost four decades ago, \cite{brown1978} reported pioneering experiments in the sputtering of water ice by H$^+$, He$^+$, C$^+$ and O$^+$ ions with MeV energies; they reported that the sputtering yield of molecules ejected from water ice per incident MeV ion was higher than predicted by contemporary theory. Brown {\it et al.} continued this sputtering work and found that at high energies electronic processes are dominant \citep{Brown1}. An electronic process represents the interaction of the projectile ions with the electrons of the target and will mostly produce ionization  and excitation \citet{Sigmund}. \citet{Cooper} used 1.6--2.5 MeV $^{19}$F ions to sputter water ice and concluded that sputtering yields are sensitive to the charge state of the ion, but independent of ice thickness and temperature below 60 K. Subsequent investigators confirmed no dependence on ice thickness or temperature at $T<60$ K \citep{Baragiola,Bar-Nun}, with the latter study finding no variation up to $T=140$ K. Furthermore, \citet{Johnson} found that water ice has a yield of 10 molecules per ion at 10 K when bombarded with 1.5 MeV He$^+$, which was in good agreement with the results obtained by \citet{brown1978}. Other experimental studies of high energy ion sputtering on water ice have reported dependencies on temperature and angle of incidence  \citep{Teolis05,Vidal,Teolis09}. The formation of CO$_2$ from 30 keV energy ion bombardment of H$_2$O ice was studied by \citet{Strazzulla03a} and \citet {Lv}

The energies and masses that make up the ion sputtering population  depends strongly on environment, whether the icy surface is within a planetary magnetosphere, within the distant solar wind or exposed solely to the galactic cosmic ray population when outside the solar heliopause \citep{Florinski13}. The ion population abundances and energies can be very different between these regions, and hence it is important to measure the sputtering yield (equivalent to the erosion rate) as a function of ion species and energy. This in turn allows modelling of the erosion rate and the amount of liberated material.
\citet{Bar-Nun} used 0.5-6 keV H$^+$ and Ne$^+$ ions to irradiate water ice and observed the sputtered species H$_2$, O$_2$ and H$_2$O.\citet{Baragiola} critically analysed data from \citet{Bar-Nun} to determine normal incidence sputtering yields due to, for example, 2-6 keV Ne$^+$ ion impact.
\citet{Christiansen} measured the sputtering yields from different ices at 78 K when bombarded with 4 keV Ar$^+$, Ne$^+$, N$^+$ and He$^+$. Finally, sputtering of water ice at 80 K by 2 keV He$^+$ and Ar$^+$ ions at normal incidence was investigated by \citet{Fama}, who also considered all the available data from previous experimental studies to propose a formula to predict the sputtering yield from water ice at any temperature by ions of any mass  with energies of up to 100 keV.

At heliocentric distances of $R_h\simeq 20-50$ AU, the solar wind plasma has cooled to $\sim 1$ eV so that the ion energy is dominated by the bulk outward flow \citep{Bagenal1997}. With a solar wind velocity of $\sim 450$ km s$^{-1}$, the kinetic energy of the ions is therefore $\simeq m_{ion}$ keV, implying that low-energy ion bombardment similar to the above experiments will be important in the Centaur region and the Kuiper Belt. Therefore to accurately model the interaction between solar wind and distant atmosphereless surfaces, it is important to test the validity of the \citet{Fama} formula for the erosion rate of water ice. In this paper we present the results of experiments on the sputtering of water ice by 4 keV singly charged ions (C$^+$, N$^+$, O$^+$ and Ar$^+$) and 4 keV doubly charged ions (C$^{2+}$, N$^{2+}$ and O$^{2+}$), and compare them to theoretical predictions and previous experimental data. Additionally, we report Temperature Programmed Desorption measurements of H$_2$O ice following sputtering by C$^{+}$ and C$^{2+}$. For all carbon ion experiments we used the isotope  $^{13}$C, but for clarity we do not designate this isotope when discussing C-ion processes.

\section{Experimental details}

The experimental equipment used in the present work was described in detail in \citet{Muntean} and will only be briefly presented here.  The full apparatus is comprised  of two main parts; a low energy ion accelerator and an ultrahigh vacuum interaction chamber. The ion source is separated from the target chamber by over 4 m of vacuum tubes with 4 stages of differential pumping. Furthermore, the pressure of source gas in the ion source is typically 10$^{-5}$ mbar. For all these reasons the level of contamination from the ion source to the target chamber is negligible. Singly and doubly charged ions with energies up to 10 keV are focused and collimated before directed onto the interaction chamber which had a base pressure of $1\times10^{-9}$ mbar. The  intensity of the ion beam is measured with a translatable Faraday Cup, which could be positioned in front of the sample. During irradiation, a 90\% transmission metal mesh is used to monitor the ion beam current.

Water ice  films were deposited onto a KBr substrate at $T=10$ K and surrounded by a gold plated radiation shield at $T=48$ K. Water ice films with thicknesses of up to 258 nm were deposited at a base pressure of $1 \times 10^{-7}$ mbar from  a nozzle via a needle valve connected to a  glass bottle containing Milli-Q water. The water was purged of dissolved gases in several cycles of freezing and heating. The film thickness was monitored with the laser interferometric method described in \citet{Muntean}; briefly, a photodiode detector monitored the intensity of $\lambda=405$ nm  laser light reflected from the ice film and KBr substrate. The angle of incidence of the light on the substrate was $45^\circ$. 

\begin{figure}
\centering
\includegraphics[width=0.5\columnwidth]{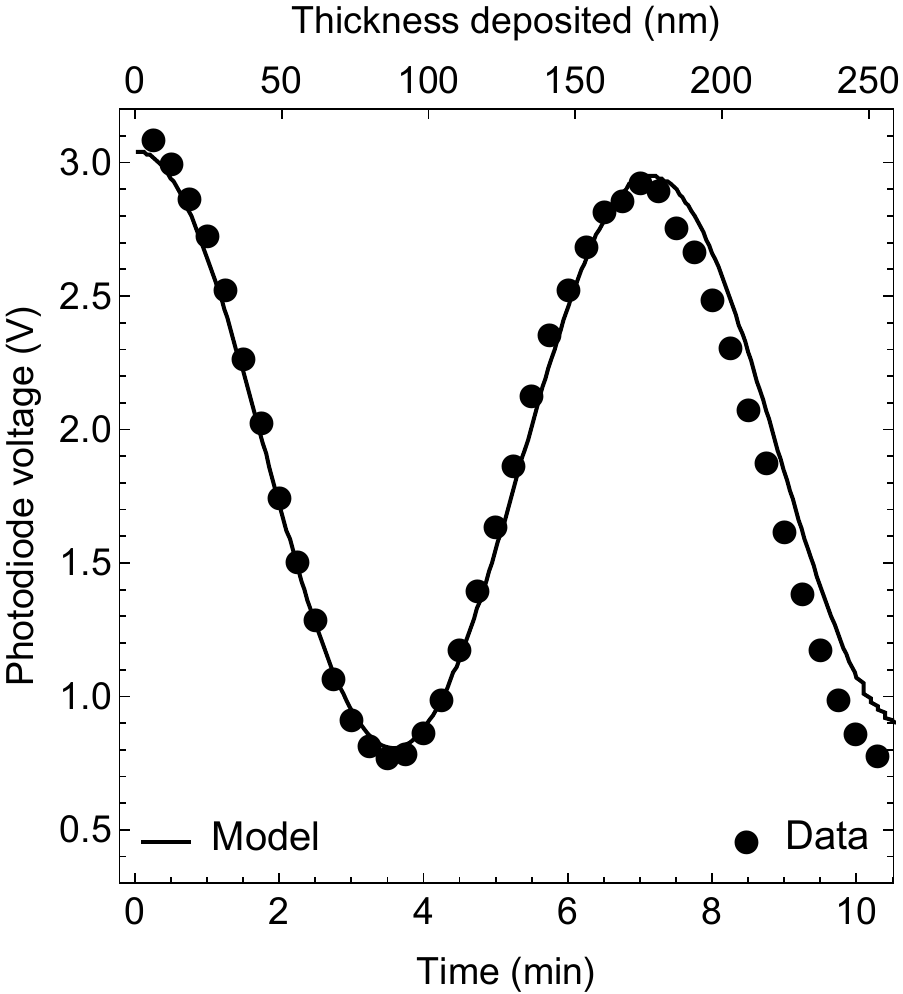}
\caption{Laser interferometer photodiode voltage (points) for experimental values and model fit (solid line) vs.\ time during the formation of water ice at 10 K from Milli-Q water at a pressure of $1.6 \times 10^{-7}$ mbar.}
\label{fig:Depositionh20}
\end{figure}

Figure \ref{fig:Depositionh20} shows the laser interferometer photodiode detector voltage as a function of time (filled circles) during the  water ice deposition process. The solid line is a model fit to the experimental data based on laws of reflection and refraction (see \citet{Muntean} for details). This enabled the refractive ice of the water ice to be determined and hence the ice density to be calculated. We obtained a best fit value of $n=1.28\pm0.05$. Using the Lorentz-Lorenz equation from \citet{Fulvio} we calculated an ice density of $\rho=0.79\pm 0.02$ g cm$^{-3}$. These results are in good agreement with values of $n=1.29\pm0.01$ and $\rho=0.82$ g cm$^{-3}$ measured by \citet{Westley} by reflection of $\lambda=435.8$ nm light from a water ice film deposited at $T=20$ K. We also note that \citet{Wood} measured the refractive index of water ice deposited at 20 K, 50 K and 80 K for 632.8 nm light and found a constant value of 1.32 for all three temperatures.

\begin{figure}
\centering
\includegraphics[width=0.5\columnwidth]{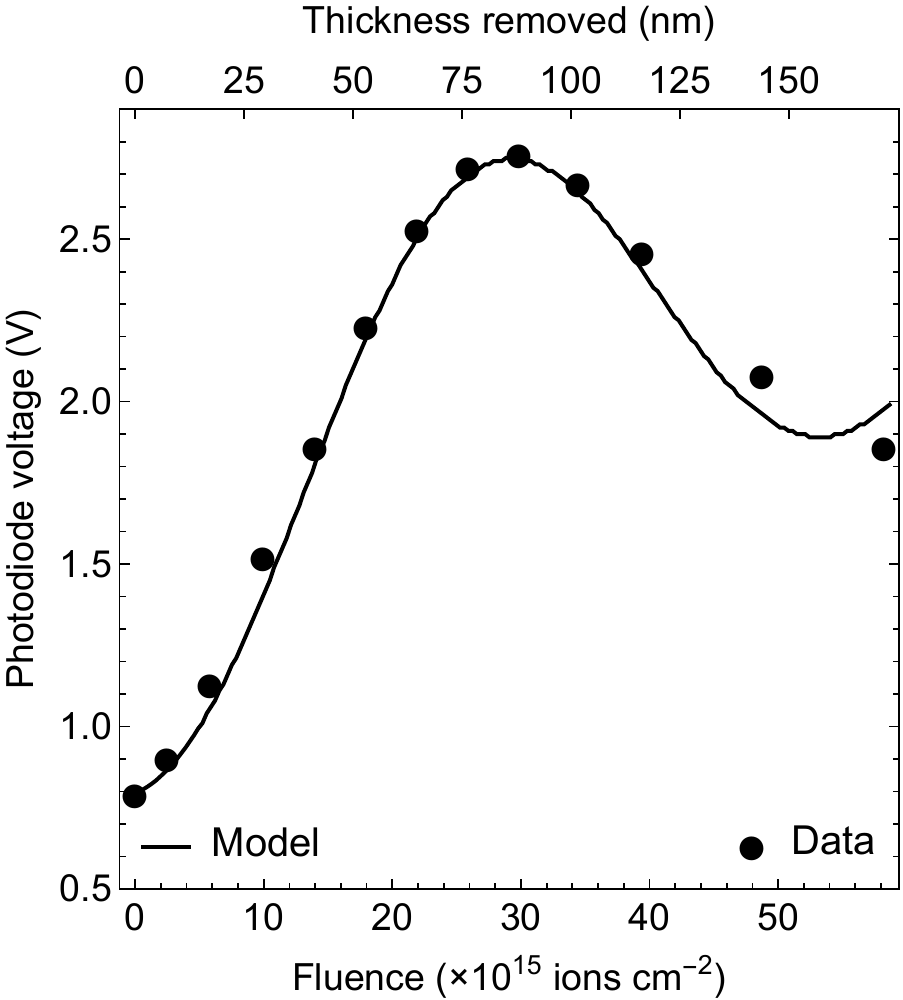}
\caption{Laser interferometric data (points) and sputtering model (solid line) for 4 keV O$^+$ on a 258.4 nm water ice film irradiated at 10 K.}
\label{fig:ThicknessO}
\end{figure}

\section{Results}

The water ice was irradiated at $T=10$ K and $45^\circ$ incidence with singly charged C$^+$, N$^+$, O$^+$ and Ar$^+$ ions and doubly charged C$^{2+}$, N$^{2+}$ and O$^{2+}$ ions. Figure \ref{fig:ThicknessO} shows the photodiode detector voltage as a function of ion dose for irradiation of an initially 258 nm thick water ice film, at 10 K by 4 keV O$^+$ ions. The solid line shown in Figure \ref{fig:ThicknessO} is a model fit using the refractive index and density determined above.  The fitted curve starts to increase before the second minimum is reached. This might be due to the irradiated ice films beginning to scatter the laser light and resulting in a loss of coherence. However, this does not affect our results as we have a good measurement of the first maximum, which is what is required for an accurate measurement of the ice thickness.
The upper horizontal scale in in Figure \ref{fig:ThicknessO} indicates the thickness of ice removed  determined by the model, which  was used to calculate the sputtering yield of $n({\rm H_2O})$/ion, {\it viz.}  the number of water molecules removed per incident ion.

\begin{figure}
\centering
\includegraphics[width=0.5\columnwidth]{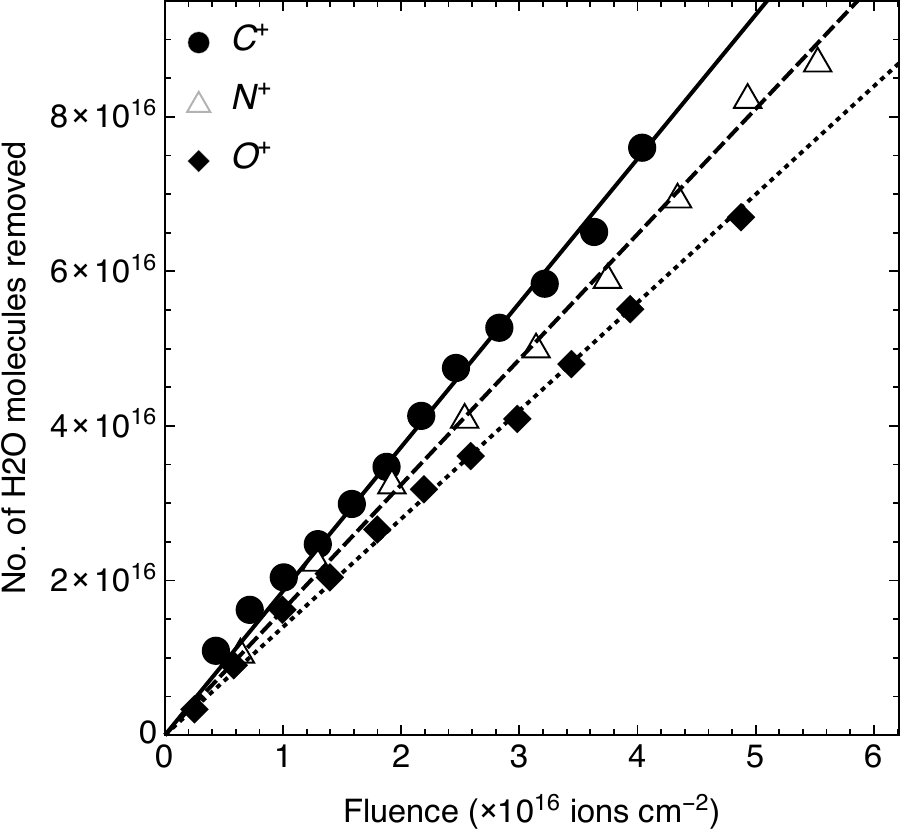}
\caption{Number of H$_2$O molecules removed by 4 keV singly charged ions on water ice at 10 K as a function of ion dose.}
\label{fig:SinglySputtering}
\end{figure}

\begin{figure}
\centering
\includegraphics[width=0.5\columnwidth]{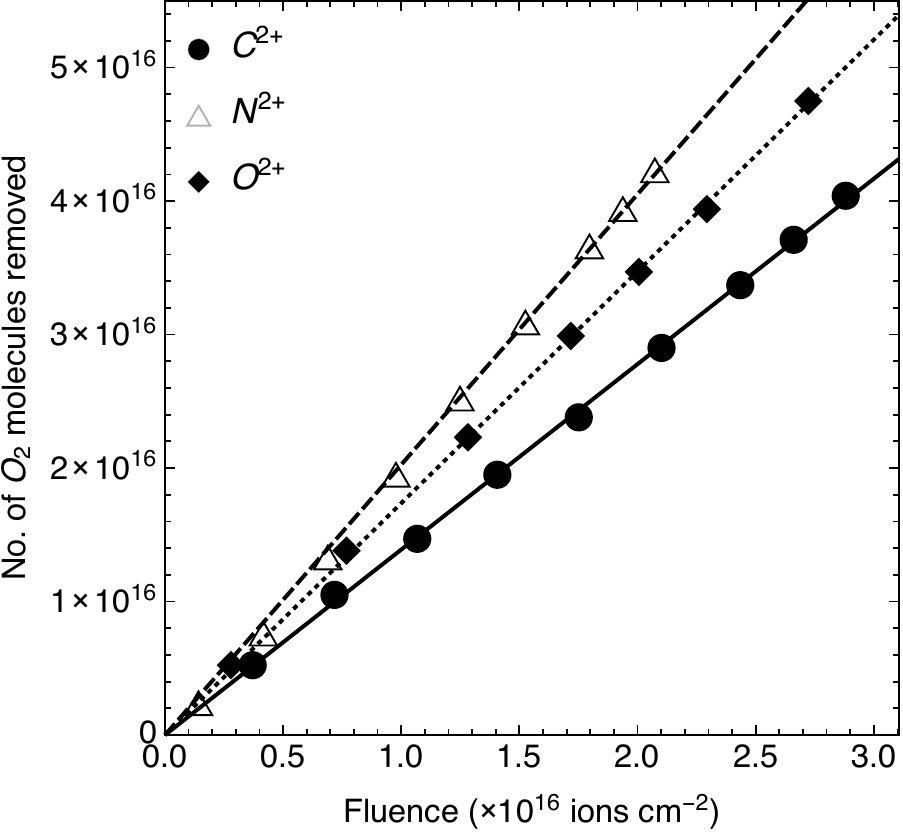}
\caption{Number of O$_2$ molecules removed by 4 keV singly charged ions on water ice at 10 K as a function of ion dose.}
\label{fig:DoublySputtering}
\end{figure}

Figure \ref{fig:SinglySputtering} and Figure \ref{fig:DoublySputtering} show the number of water molecules removed as a function of  fluence for 4 keV singly charged C$^+$, N$^+$ and O$^+$  and for 4 keV doubly charged C$^{2+}$, N$^{2+}$ and O$^{2+}$ ions respectively. The ion beam intensity varied both for individual ions and range of ions studied, thus we express it in fluence rather then flux. Typically ion fluxes  averages were about
$1.8\times10^{12}$ ions/cm$^2$/s for doubly charged ions and $9\times 10^{12}$ ions/cm$^2$/s for singly charged ions.
From the slopes of  linear regressions fitted to the experimental data and the measured ice density and refractive index, the sputtering yields for each ion species were calculated and are shown in Table \ref{table:1}. Taking our uncertainty in the fitted refractive index and propagating it through this analysis gives a total error budget of up to 20\%. We note that the fitting error in Figures 3 and 4 is $\sim0.1$\%.

\begin{table}
  \caption{Experimental and theoretical values for sputtering yield of water ice (molecules per incident ion) at 10 K by 4 keV singly and doubly charged ions. The uncertainties shown are the  total experimental errors.} % title of Table
\label{table:1} % is used to refer this table in the text
\centering % used for centering table
\begin{tabular}{ccccc} % centered columns (4 columns)
\hline \hline % inserts double horizontal lines
\noalign{\smallskip}
        & $n$  & $\rho$ (g cm$^{-3}$) &  Y(H$_2$O) Expt. & Y(H$_2$O) Theory \\
         \hline
         \noalign{\smallskip}
 C$^+$    & 1.282 & 0.799 & $10.5\pm2.1$   & 7.3 	\\
 N$^+$    & " & " & $9.2\pm1.8$  & 8.8 	\\
 O$^+$    & " & " & $7.9\pm1.6$ & 9.7 	\\
 Ar$^+$   & " & " & $19.8\pm4.0$   & 	17.1	\\
 \hline
 \noalign{\smallskip}
 C$^{2+}$ & " & " & $7.8\pm1.6$  & $\cdots$ \\
 N$^{2+}$ & " & " & $11.4\pm2.3$  & $\cdots$ \\
 O$^{2+}$ & " & " & $9.8\pm1.8$ & $\cdots$ \\
\hline %inserts single line
\end{tabular}
\end{table}

\begin{figure}
\centering
\includegraphics[width=0.5\columnwidth]{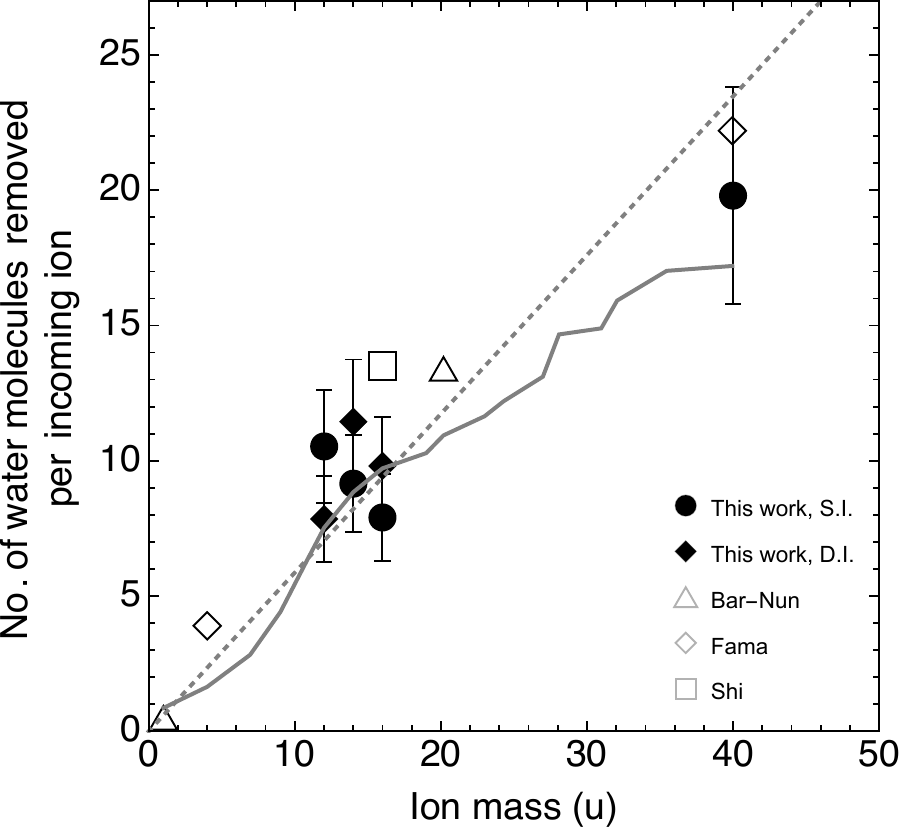}
\caption{Sputtering yield per incoming ion as a function of ion mass. Our experimental results are marked as circles (filled symbols indicate singly ionised species and open symbols correspond to doubly ionised species.) Other symbols indicate experimental data by other authors \citep{Bar-Nun,Fama,Shi}. The solid line indicates the theoretical yield as a function of projectile mass, and the dashed line shows a simple linear fit (slope $0.59\pm0.03$) to all experimental data.}
\label{fig:YieldPerIon}
\end{figure}

Our experimental sputter yields for singly charged ions have been compared with theoretical values predicted by the model developed by \citet{Fama}, given by the equation below. 

\begin{equation}
Y\left( E,m_1,Z_1,\theta,T \right) = \frac{1}{U_0}\left(\frac{3}{4\,\pi^2\,C_0}\alpha S_n + \eta\,S_e^2 \right) \times \left( 1 + \frac{Y_1}{Y_0}\exp{\frac{-E_0}{k\,T}} \right)\,\cos^{-f}\theta
\end{equation}

Fam\'a's model predicts the sputtering yield as a result of ion irradiation of water ice at temperatures $T$ from 10 K to 140 K, with ion energies $E$ up to 100 keV and projectile ions with mass $m_1$, atomic number $Z_1$ and incident angle $\theta$. Our sputtering  experiments were carried at a constant temperature (10 K) and an ion energy of 4 keV which means $\frac{Y_1}{Y_0}\exp{\frac{-E_0}{k\,T}} \simeq 0$. Also, the incidence angle, $\theta$, was 45 degrees for all the experiments. $S_n$ and $S_e$ are the nuclear and electronic stopping cross-sections respectively, which were calculated using the standard Stopping Range of Ions in Matter (SRIM\footnote{http://www.srim.org/}) software. $f$ denotes the power of the angular scattering function, $\alpha$ is a function of the mass ratio between the target and the projectile atomic masses and $\eta$ is a function of the atomic number of the projectile. The empirical formulae used for obtaining the energy independent values of $\eta$, $\alpha$ and  $f$ are explained in detail in \citet{Fama}. $U_0=0.45$~eV is the binding energy, and C$_0$=1.81 \AA$^2$ is a constant which is related to the differential cross section for elastic scattering in the binary collision approximation  \citep{Sigmund}.

Table \ref{table:1} summarizes the experimental and the predicted values for the sputtering yield of water ice when bombarded with singly and doubly charged ions. It can be clearly seen that the sputtering yield for both singly and doubly charged ions are the same within the experimental error. In Figure \ref{fig:YieldPerIon} we plot our experimental values for the sputtering yield together with previous measurements in the literature. Our experimentally determined values for the sputtering yields are consistent with previous reports. For example, \citet{Fama} measured a sputtering yield for Ar$^+$ at normal incidence of 12 molecules/incident ion. This equates to 21.8 molecules/ion at 45 degrees incidence, consistent with our measured value of  $19.8\pm4.0$ molecules/ion.
Recently, \citet{Galli} have measured the sputtering yield of porous NaCl water ice at 90 K for a range of ions with energies between 1 and 30 keV ; the values obtained are consistent with all the data shown in Figure \ref{fig:YieldPerIon}. The data of Galli et al. are not plotted however, as the uncertainties of their measurements extend beyond the range of the ordinate. In Figure \ref{fig:YieldPerIon} we also show the predicted values from the \citet{Fama} theoretical model and find good agreement.  We note that there is a slight tendency for the model to underestimate the experimental yields, the reason for this is unclear. 

In addition, temperature programmed desorption (TPD) measurements  were carried out after the irradiation by $^{13}$C$^+$ and $^{13}$C$^{2+}$ with a constant  ramp rate of 1 K/ min. As can be seen from Figure 6 and Table \ref{table:2}, irradiation by C$^+$ produced more CO$_2$ than irradiation by C$^{2+}$, with a yield ratio of $2.1\pm0.2$.
This situation was reversed in the case of CO as shown in figure 7, where the ratio of yields from C$^+$ and  C$^{2+}$ was $0.4\pm0.02$.
For C$^+$ irradiation, the production ratio of CO/CO$_2 \simeq 27$. For C$^{2+}$ irradiation, the production ratio of CO/CO$_2 \simeq 130$

\begin{figure}
\centering
\includegraphics[width=0.5\columnwidth]{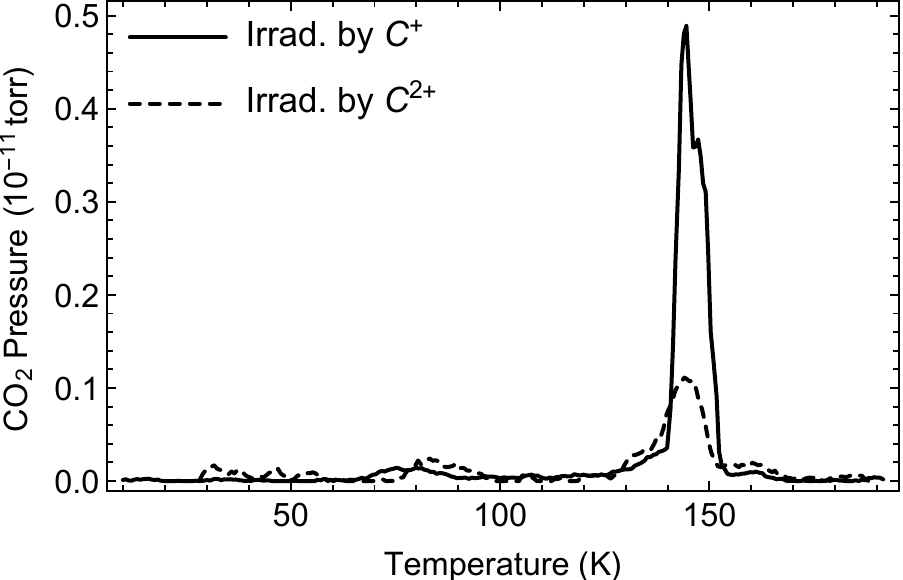}
\caption{TPD profile for CO$_2$ produced through irradiation of H$_2$O ice by C$^+$ and C$^{2+}$ (total dose $10^{16}$ ions in each case). The dominant TPD peak occurs at $T=144.4\pm0.1$ K, while two lower peaks are found at $T=84.7\pm0.4$ K and $T=160.0\pm0.3$ K.}
\label{fig:TPDCO2}
\end{figure}

\begin{figure}
\centering
\includegraphics[width=0.5\columnwidth]{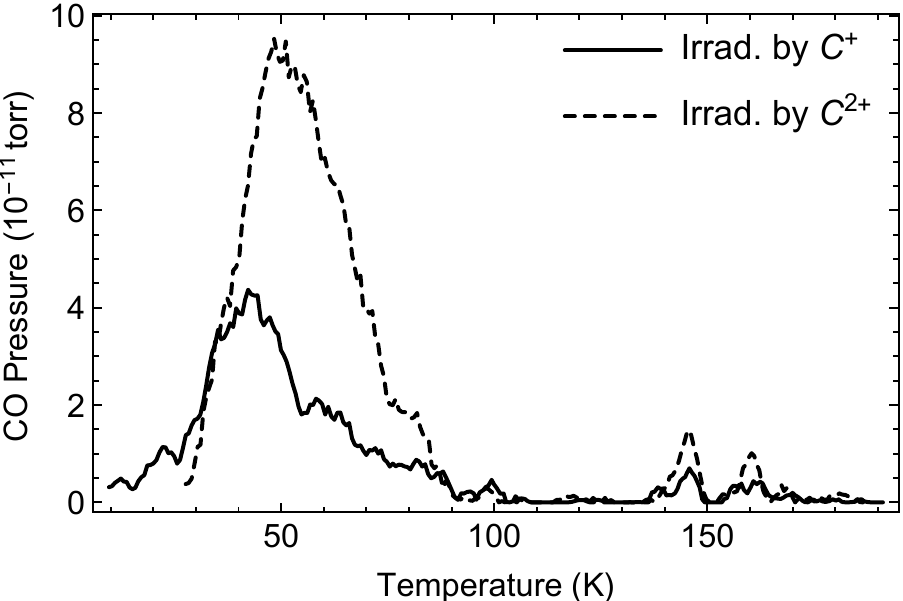}
\caption{TPD profile for CO produced through irradiation of H$_2$O ice by C$^+$ and C$^{2+}$ (total dose $10^{16}$ ions in each case). The dominant TPD peak occurs at $T=42.5$ K after irradiation with C$^+$ and at $T=50.3$ K after irradiation with C$^{2+}$, while two lower peaks are found near $T=145$ K and $T=160.0$ K in both cases.}
\label{fig:TPDCO}
\end{figure}

\begin{table}
\caption{Temperature programmed desorption temperature peaks seen in Figures \ref{fig:TPDCO2} and \ref{fig:TPDCO}  Peak properties (central temperature, $T$, and area under the curve, $I$, in units of K Torr) derived from Gaussian fits to the data.} % title of Table
\label{table:2} % is used to refer this table in the text
\centering % used for centering table
\begin{tabular}{crrrr} % centered columns (4 columns)
\hline \hline % inserts double horizontal lines
\noalign{\smallskip}
       & \multicolumn{2}{c}{CO} &  \multicolumn{2}{c}{CO$_2$} \\
Ion &  $T$~(K) &  $I/10^{-11}$ &  $T$~(K) & $I/10^{-11}$      \\
\hline
\noalign{\smallskip}
C$^+$    &  $42.5\pm0.1$  & $102.0\pm3.5$ & $78.0\pm0.2$  &  $0.25\pm0.02$ \\
C$^+$    &  $145.5\pm0.1$ & $3.30\pm0.39$ & $145.5\pm0.1$ &  $3.68\pm0.29$  \\
C$^+$    &  $160.0\pm0.5$ & $3.72\pm0.86$ & $160.0\pm0.2$ &  $0.08\pm0.01$  \\
\hline
\noalign{\smallskip}
C$^{2+}$ &  $50.3\pm0.3$  & $239.3\pm12.9$ & $84.7\pm0.4$  &  $0.35\pm0.04$  \\
C$^{2+}$ &  $145.1\pm0.2$ & $8.32\pm1.02$  & $144.4\pm0.1$ &  $1.37\pm0.08$  \\
C$^{2+}$ &  $160.0\pm0.1$ & $4.76\pm0.31$  & $160.0\pm0.3$ &  $0.22\pm0.04$  \\
\hline %inserts single line
\end{tabular}
\end{table}

\section{Discussion}

\subsection{Outer solar system sputtering rates}

Our measured sputter yield can be used to estimate at what rate H$_2$O ice is removed from the surface of an outer solar system object due to irradiation by ions in the solar wind (SW). Taking a SW ion density of $n_i=4.5$ cm$^{-3}$ and velocity $v_i=400$ km s$^{-1}$ at 1 AU \citep{Johnson} we calculate a total ion flux at a nominal distance of $a=42$ AU of $v_i\,n_i/a^2=10^5$ ions cm$^{-2}$ s$^{-1}$. Table \ref{table:3} lists the relative abundances used here, calculated using the Genesis measurement of He/H fluence in the SW \citep{Reisenfeld} and the values for C/He, N/He, and O/He from the Ulysses mission \citep{VonSteiger}. We combine these measurements to give the predicted the fluxes at 42 AU for ions of H, He, C, N and O. Our O$^+$ irradiation experiment indicates that sputtering erodes 86 nm of H$_2$O ice for a fluence of $4.5\times10^{15}$ O$^+$ ions (Figure \ref{fig:ThicknessO}). As seen in Figure \ref{fig:YieldPerIon}, this sputter yield scales as $0.59\,m_i$ where $m_i$ is the mass of the ion in amu. Using these figures we calculate in Table \ref{table:3} the expected erosion rate for H$_2$O at 42 AU due to sputtering by SW ions. 

We find that proton erosion dominates, agreeing with expectations. It is also clear that sputtering will significantly erode and alter the optically active surface layers of ice in the Trans-Neptunian region on extremely short timescales relative to dynamical lifetimes of observed objects  of $\geq 4.5\times10^9$ years \citep{Duncan95,Morbidelli08}. Other laboratory experiments have investigated  the reddening of ices due to ion irradiation. 
According to the measurements of \citet{Strazzulla03b} and \citet{Brunetto08}, it optimistically takes $\sim10^9$ years to produce a 1 micron-deep reddened irradiation mantle.
Sputtering by solar wind ions to a comparable depth is three orders of magnitude faster. This leads us to two possibilities: Either sputtering does not remove the red irradiation mantle hypothesised to exist on outer solar system icy bodies; or the material is naturally red and reddening by ion irradiation is not a dominant effect.  

We point out, however, that UV irradiation is an alternative pathway to the production of complex molecules \citep{bernstein97}, a process that may occur on significantly faster timescales than the ion-induced changes. Also, an initial composition of complex hydrocarbons can give rise to significant reddening on dramatically shorter timescales \citep{Kanuchova12}. As such, we suggest an appropriate amount of caution when interpreting reddening timescales implied by laboratory measurements.

% * <a.fitzsimmons@qub.ac.uk> 2015-10-30T17:51:59.960Z:
%
% ^.
% * <a.fitzsimmons@qub.ac.uk> 2015-10-30T17:51:52.436Z:
%
% ^.
% * <a.fitzsimmons@qub.ac.uk> 2015-10-30T17:51:51.361Z:
%
% ^.

Sputtering of surface ices may be even more important for water-ice rich Centaur objects evolving inwards from the trans-Neptunian region into the giant planet region.
Many Centaurs have spectroscopically confirmed H$_2$O ice absorption bands on their surfaces \citep{Barkume08, Barucci11}.
Lifetimes of Centaurs typically lie in the range
10--90 Myr \citep{Tiscareno03, Volk13}. At $\sim10$ AU the erosion rate will be $(42/10)^2\sim 20$ times faster, so the solar wind will erode $\simeq 0.1-1$ mm of H$_2$O ice on the surface of a Centaur during its dynamical lifetime.
CO and CO$_2$ ices could also be present and will sublimate at these distances, although the observed activity in a small number of Centaurs may be due to this or due to crystallization of amorphous H$_2$O ice \citep{Jewitt09,Lepoutre12}. Our results show that even without sublimation or crystallization, pure H$_2$O ice will  not survive unchanged. Therefore we conclude that the optically active surfaces of all ice-rich bodies beyond Jupiter will be significantly modified.

\begin{table}
\caption{Solar wind properties and derived H$_2$O sputtering rate.} % title of Table
\label{table:3} % is used to refer this table in the text
\centering % used for centering table
\begin{tabular}{ccccc} % centered columns (4 columns)
\hline \hline % inserts double horizontal lines
\noalign{\smallskip}
SW Ion & Relative & Flux at 42 AU & Sputtering rate\\
& Abundance & [cm$^{-2}$ Myr$^{-1}$] & [nm Myr$^{-1}$] \\
\hline
\noalign{\smallskip}
H   & $9.54\times10^{-1}$ & $3.1\times10^{18}$ & $3.20\times10^3$ \\
He  & $4.58\times10^{-2}$ & $1.5\times10^{17}$ & $614$ \\
C   & $3.39\times10^{-4}$ & $1.1\times10^{15}$ & $13.6$ \\
N   & $3.98\times10^{-5}$ & $1.3\times10^{14}$ & $1.87$ \\
O   & $5.04\times10^{-4}$ & $1.6\times10^{15}$ & $27.0$ \\
Fe  & $5.54\times10^{-5}$ & $1.8\times10^{14}$ & $10.4$ \\
\hline %inserts single line
\end{tabular}
\end{table}

\subsection{Formation of CO and CO$_2$ by carbon ions}

Our TPD results can be interpreted qualitatively in terms of CO and CO$_2$ forming in the H$_2$O ice as a result of irradiation by C$^+$ and C$^{2+}$. It is interesting to compare these results with TPD measurements of \citet{Collings} who investigated separate water ices  with CO and with CO$_2$ on the surface and with CO and with CO$_2$ in the bulk. Surface samples were prepared by deposition on top of water ice and samples with bulk CO and CO$_2$ were prepared by co-deposition of either CO or CO$_2$ with H$_2$O. In our experiments the majority of CO$_2$ desorption occurs at $T\sim145$ K, $92 \%\pm 10\% $ in the case of C$^{+}$ irradiation and $71\%\pm 5 $\% in the case of C$^{2+}$ irradiation. This 145 K release was observed by \citet{Collings}, who attribute this in their experiments to the volcano desorption of trapped CO$_2$ co-deposited with H$_2$O. 

The release of CO$_2$ at around $T\sim160$ K for both singly and doubly charged ion irradiation corresponds to co-desorption of trapped molecules with H$_2$O. However significantly less CO$_2$ was released at this temperature, $2\pm 0.3\% $\% for singly charged ions and $11\%\pm 2 $\% for doubly charged ions. This differs from the results presented by \citet{Collings}, who found roughly equal amounts desorbed at $T\sim145$ K and $T\sim160$ K when the CO$_2$ was co-deposited with H$_2$O.

For CO, our TPD results show major broad peaks at lower temperatures. In C$^+$ irradiation  $93\%\pm 4 $\% of the CO desorbs at $T=42$ K, whereas in C$^{2+}$ irradiation $92\%\pm 2 $\% desorbs at $T=50$ K. The peaks at $T=145$ K and $T=160$ K  that correspond to volcano desorption and co-desorption respectively are minor features. Similar desorption features were observed by \citet{Collings} for CO both deposited onto a pre-adsorbed H$_2$O film, and co-deposited with H$_2$O. They also observed broad desorption peaks at low temperatures, but with similar amounts of CO released at the higher temperatures.
The dominance of the low temperature peaks in the CO production, particularly in the case of C$^{2+}$, suggests that charge dependent dissociation processes are occurring at or close to the surface of the water ice. In the case of CO$_2$, the majority is trapped below the surface layers within the H$_2$O matrix. It is interesting to note that different chemical reactions due to irradiation have been observed in layered H$_2$O/CO/H$_2$O films in which CO layers were buried under amorphous solid water films of different thicknesses and then irradiated with 100 eV electrons \cite{petrik2014}. They found that the products formed from the oxidation and reduction/hydrogenation of CO were different for buried depths of up to 50 monolayers compared to products of CO at depths of more than 50 monolayers. Our work suggests a depth dependent effect in CO and CO$_2$ production.

Our yields in Table 2. imply that the mechanism for the formation of CO$_2$ is more efficiently driven by C$^+$ irradiation by approximately a factor of 2 compared to C$^{2+}$ irradiation. A similar result was found by  \citet{Dawes}, who  irradiated  water ice at 30 K and 90 K with 4 keV C$^+$ and C$^{2+}$ and measured  CO$_2$ formation as a function of ion dose. They
concluded that that the CO$_2$ yield for singly charged ions is greater than for doubly charged ions, in qualitative agreement with our findings. 

A significant finding from our TPD measurements was the observed large production rate of CO relative to CO$_2$.
4keV C ions have a  maximum penetration depth in water ice of 500 \AA with a peak range of 236 \AA. If in our case we take the peak range and the thickness removed (1700 \AA), this means only 14\% of  the  C$^{2+}$ ions that can remain implanted to form CO and CO$_2$. Therefore, for a total fluence  for C$^{2+}$ of $2.9\times10^{16}$ ions/cm$^2$, the maximum amount of retained ions becomes $2.1\times10^{15} $ions/cm$^2$. For a production yield of $\sim0.5$, a beam area of $1.77\times10^{-1} $cm$^2$ and a measured abundance of CO/CO$_2$=130 we have calculated an  upper limit of the CO$_2$ produced  to be $1.4\times10^{12}$ molecules. In their previous study of 4 keV C-ion bombardment of H$_2$O ice, \citet{Dawes} did not observe {\em any} CO. However, we note that their lowest experimental temperature was 30 K. This is where the  sublimation of CO starts according to both our TPD measurements (Figure 7) and those of \citet{Collings}. Therefore we propose that as the CO formed preferentially at the surface, it was immediately sublimated during the \citet{Dawes} study, thereby accounting for the lack of subsequent spectral FTIR signatures.

Finally, it is interesting to note that in a 30 keV study at 10 K by \citet{Lv} and \citet{Strazzulla03a} CO was not observed. This intriguing result may imply that CO formation by carbon ion bombardment of low temperature ice is highly energy dependent, and deserves further study.

\section{Conclusions}

We have measured the sputtering yield from irradiated ice at 10 K by 4 keV singly charged (C$^+$, N$^+$, O$^+$ and Ar$^+$) and doubly charged ions (C$^{2+}$, N$^{2+}$ and O$^{2+}$). In the case of singly charged ions, the sputtering yields have been compared with a theoretical model from \citet{Fama} and overall our measurements are in line with previous results and the model`s predictions. Within experimental uncertainties, the sputtering yield of water ice does not depend on whether the projectiles are singly or doubly charged ions. We confirm that Solar wind sputtering of fresh ice surfaces in the Kuiper belt and Centaur region can erode H$_2$O on dynamically relevent timescales.

Temperature programmed desorption showed that CO formation dominates over CO$_2$ formation. However $^{13}$C$^{2+}$ is relatively more efficient in forming  $^{13}$CO while $^{13}$C$^+$ is relatively more efficient in forming $^{13}$CO$_2$. The majority of the $^{13}$CO desorbs at $T\simeq42$ K which suggests that this species is produced close to the surface of the water ice, whereas the majority of the $^{13}$CO$_2$ desorbs at $T\simeq145$ K and indicates that CO$_2$ is formed below the surface.

\section*{Acknowledgements}
% Entry for the table of contents, for this guide only
\addcontentsline{toc}{section}{Acknowledgements}

 We thank M. Fam{\'a} for  useful comments and suggestions.
      We thank the Leverhulme Trust for support through a Research Project Grant RPG-2013-389. 
      PL is grateful for support from the Royal Society in the form of a Newton Fellowship. 
      Part of this work was supported by the European Commission's 7th Framework Programme under Grant Agreement No. 238258.
%%%%%%%%%%%%%%%%%%%%%%%%%%%%%%%%%%%%%%%%%%%%%%%%%%



%\bibliographystyle{mnras}

\bibliographystyle{mnras}
\bibliography{references}

\begin{thebibliography}{}
\makeatletter
\relax
\def\mn@urlcharsother{\let\do\@makeother \do\$\do\&\do\#\do\^\do\_\do\%\do\~}
\def\mn@doi{\begingroup\mn@urlcharsother \@ifnextchar [ {\mn@doi@}
  {\mn@doi@[]}}
\def\mn@doi@[#1]#2{\def\@tempa{#1}\ifx\@tempa\@empty \href
  {http://dx.doi.org/#2} {doi:#2}\else \href {http://dx.doi.org/#2} {#1}\fi
  \endgroup}
\def\mn@eprint#1#2{\mn@eprint@#1:#2::\@nil}
\def\mn@eprint@arXiv#1{\href {http://arxiv.org/abs/#1} {{\tt arXiv:#1}}}
\def\mn@eprint@dblp#1{\href {http://dblp.uni-trier.de/rec/bibtex/#1.xml}
  {dblp:#1}}
\def\mn@eprint@#1:#2:#3:#4\@nil{\def\@tempa {#1}\def\@tempb {#2}\def\@tempc
  {#3}\ifx \@tempc \@empty \let \@tempc \@tempb \let \@tempb \@tempa \fi \ifx
  \@tempb \@empty \def\@tempb {arXiv}\fi \@ifundefined
  {mn@eprint@\@tempb}{\@tempb:\@tempc}{\expandafter \expandafter \csname
  mn@eprint@\@tempb\endcsname \expandafter{\@tempc}}}

\bibitem[\protect\citeauthoryear{{Bagenal}, {Cravens}, {Luhmann}, {McNutt}  \&
  {Cheng}}{{Bagenal} et~al.}{1997}]{Bagenal1997}
{Bagenal} F.,  {Cravens} T.~E.,  {Luhmann} J.~G.,  {McNutt} Jr. R.~L.,
  {Cheng} A.~F.,  1997, in {Stern} S.~A.,  {Tholen} D.~J.,  eds, , Pluto and
  Charon.
Oxford University Press, Oxford, p.~523

\bibitem[\protect\citeauthoryear{Bar-Nun, Herman, Rappaport  \& Mekler}{Bar-Nun
  et~al.}{1985}]{Bar-Nun}
Bar-Nun A.,  Herman G.,  Rappaport M.,   Mekler Y.,  1985, Surface science,
  150, 143

\bibitem[\protect\citeauthoryear{Baragiola, Vidal, Svendsen, Schou, Shi, Bahr
  \& Atteberrry}{Baragiola et~al.}{2003}]{Baragiola}
Baragiola R.,  Vidal R.,  Svendsen W.,  Schou J.,  Shi M.,  Bahr D.,
  Atteberrry C.,  2003, Nuclear Instruments and Methods in Physics Research
  Section B: Beam Interactions with Materials and Atoms, 209, 294

\bibitem[\protect\citeauthoryear{{Barkume}, {Brown}  \& {Schaller}}{{Barkume}
  et~al.}{2008}]{Barkume08}
{Barkume} K.~M.,  {Brown} M.~E.,   {Schaller} E.~L.,  2008, \mn@doi [\aj]
  {10.1088/0004-6256/135/1/55}, \href
  {http://adsabs.harvard.edu/abs/2008AJ....135...55B} {135, 55}

\bibitem[\protect\citeauthoryear{{Barucci}, {Alvarez-Candal}, {Merlin},
  {Belskaya}, {de Bergh}, {Perna}, {DeMeo}  \& {Fornasier}}{{Barucci}
  et~al.}{2011}]{Barucci11}
{Barucci} M.~A.,  {Alvarez-Candal} A.,  {Merlin} F.,  {Belskaya} I.~N.,  {de
  Bergh} C.,  {Perna} D.,  {DeMeo} F.,   {Fornasier} S.,  2011, \mn@doi
  [\icarus] {10.1016/j.icarus.2011.04.019}, \href
  {http://adsabs.harvard.edu/abs/2011Icar..214..297B} {214, 297}

\bibitem[\protect\citeauthoryear{{Bernstein}, {Allamandola}  \&
  {Sandford}}{{Bernstein} et~al.}{1997}]{bernstein97}
{Bernstein} M.~P.,  {Allamandola} L.~J.,   {Sandford} S.~A.,  1997, \mn@doi
  [Advances in Space Research] {10.1016/S0273-1177(97)00340-2}, \href
  {http://adsabs.harvard.edu/abs/1997AdSpR..19..991B} {19, 991}

\bibitem[\protect\citeauthoryear{{Brown} \& {Hill}}{{Brown} \&
  {Hill}}{1996}]{Brown1996}
{Brown} M.~E.,  {Hill} R.~E.,  1996, \mn@doi [\nat] {10.1038/380229a0}, \href
  {http://adsabs.harvard.edu/abs/1996Natur.380..229B} {380, 229}

\bibitem[\protect\citeauthoryear{Brown, Lanzerotti, Poate  \&
  Augustyniak}{Brown et~al.}{1978}]{brown1978}
Brown W.,  Lanzerotti L.,  Poate J.,   Augustyniak W.,  1978, Physical Review
  Letters, 40, 1027

\bibitem[\protect\citeauthoryear{Brown, Augustyniak, Lanzerotti, Johnson  \&
  Evatt}{Brown et~al.}{1980}]{Brown1}
Brown W.,  Augustyniak W.,  Lanzerotti L.,  Johnson R.,   Evatt R.,  1980,
  Physical Review Letters, 45, 1632

\bibitem[\protect\citeauthoryear{{Brown}, {Schaller}  \& {Fraser}}{{Brown}
  et~al.}{2012}]{Brown2012}
{Brown} M.~E.,  {Schaller} E.~L.,   {Fraser} W.~C.,  2012, \mn@doi [\aj]
  {10.1088/0004-6256/143/6/146}, \href
  {http://adsabs.harvard.edu/abs/2012AJ....143..146B} {143, 146}

\bibitem[\protect\citeauthoryear{{Brunetto} \& {Roush}}{{Brunetto} \&
  {Roush}}{2008}]{Brunetto08}
{Brunetto} R.,  {Roush} T.~L.,  2008, \mn@doi [\aap]
  {10.1051/0004-6361:20078889}, \href
  {http://adsabs.harvard.edu/abs/2008A%26A...481..879B} {481, 879}

\bibitem[\protect\citeauthoryear{{Brunetto}, {Barucci}, {Dotto}  \&
  {Strazzulla}}{{Brunetto} et~al.}{2006}]{Brunetto2006}
{Brunetto} R.,  {Barucci} M.~A.,  {Dotto} E.,   {Strazzulla} G.,  2006, \mn@doi
  [\apj] {10.1086/503359}, \href
  {http://adsabs.harvard.edu/abs/2006ApJ...644..646B} {644, 646}

\bibitem[\protect\citeauthoryear{Christiansen, Carpini  \& Tsong}{Christiansen
  et~al.}{1986}]{Christiansen}
Christiansen J.,  Carpini D.~D.,   Tsong I.,  1986, Nuclear Instruments and
  Methods in Physics Research Section B: Beam Interactions with Materials and
  Atoms, 15, 218

\bibitem[\protect\citeauthoryear{{Collings}, {Anderson}, {Chen}, {Dever},
  {Viti}, {Williams}  \& {McCoustra}}{{Collings} et~al.}{2004}]{Collings}
{Collings} M.~P.,  {Anderson} M.~A.,  {Chen} R.,  {Dever} J.~W.,  {Viti} S.,
  {Williams} D.~A.,   {McCoustra} M.~R.~S.,  2004, \mn@doi [\mnras]
  {10.1111/j.1365-2966.2004.08272.x}, \href
  {http://adsabs.harvard.edu/abs/2004MNRAS.354.1133C} {354, 1133}

\bibitem[\protect\citeauthoryear{Cooper \& Tombrello}{Cooper \&
  Tombrello}{1984}]{Cooper}
Cooper B.,  Tombrello T.,  1984, Radiation effects, 80, 203

\bibitem[\protect\citeauthoryear{{Dalton}, {Cruikshank}, {Stephan}, {McCord},
  {Coustenis}, {Carlson}  \& {Coradini}}{{Dalton} et~al.}{2010}]{Dalton}
{Dalton} J.~B.,  {Cruikshank} D.~P.,  {Stephan} K.,  {McCord} T.~B.,
  {Coustenis} A.,  {Carlson} R.~W.,   {Coradini} A.,  2010, \mn@doi [\ssr]
  {10.1007/s11214-010-9665-8}, \href
  {http://adsabs.harvard.edu/abs/2010SSRv..153..113D} {153, 113}

\bibitem[\protect\citeauthoryear{Dawes, Hunniford, Holtom, Mukerji, McCullough
  \& Mason}{Dawes et~al.}{2007}]{Dawes}
Dawes A.,  Hunniford A.,  Holtom P.~D.,  Mukerji R.~J.,  McCullough R.~W.,
  Mason N.~J.,  2007, Physical Chemistry Chemical Physics, 9, 2886

\bibitem[\protect\citeauthoryear{{Duncan}, {Levison}  \& {Budd}}{{Duncan}
  et~al.}{1995}]{Duncan95}
{Duncan} M.~J.,  {Levison} H.~F.,   {Budd} S.~M.,  1995, \mn@doi [\aj]
  {10.1086/117748}, \href {http://adsabs.harvard.edu/abs/1995AJ....110.3073D}
  {110, 3073}

\bibitem[\protect\citeauthoryear{{Fam{\'a}}, {Shi}  \& {Baragiola}}{{Fam{\'a}}
  et~al.}{2008}]{Fama}
{Fam{\'a}} M.,  {Shi} J.,   {Baragiola} R.~A.,  2008, \mn@doi [Surface Science]
  {10.1016/j.susc.2007.10.002}, \href
  {http://www.researchgate.net/publication/229133343_Sputtering_of_ice_by_low-energy_ions}
  {602, 156}

\bibitem[\protect\citeauthoryear{{Florinski}, {Jokipii}, {Alouani-Bibi}  \& {le
  Roux}}{{Florinski} et~al.}{2013}]{Florinski13}
{Florinski} V.,  {Jokipii} J.~R.,  {Alouani-Bibi} F.,   {le Roux} J.~A.,  2013,
  \mn@doi [\apjl] {10.1088/2041-8205/776/2/L37}, \href
  {http://adsabs.harvard.edu/abs/2013ApJ...776L..37F} {776, L37}

\bibitem[\protect\citeauthoryear{Fulvio, Sivaraman, Baratta, Palumbo  \&
  Mason}{Fulvio et~al.}{2009}]{Fulvio}
Fulvio D.,  Sivaraman B.,  Baratta G.,  Palumbo M.,   Mason N.,  2009,
  Spectrochimica Acta Part A: Molecular and Biomolecular Spectroscopy, 72, 1007

\bibitem[\protect\citeauthoryear{Galli et~al.,}{Galli et~al.}{2015}]{Galli}
Galli A.,  et~al., 2015, arXiv preprint arXiv:1509.04008

\bibitem[\protect\citeauthoryear{{Guilbert-Lepoutre}}{{Guilbert-Lepoutre}}{2012}]{Lepoutre12}
{Guilbert-Lepoutre} A.,  2012, \mn@doi [\aj] {10.1088/0004-6256/144/4/97},
  \href {http://adsabs.harvard.edu/abs/2012AJ....144...97G} {144, 97}

\bibitem[\protect\citeauthoryear{{Guilbert}, {Alvarez-Candal}, {Merlin},
  {Barucci}, {Dumas}, {de Bergh}  \& {Delsanti}}{{Guilbert}
  et~al.}{2009}]{Guilbert2009}
{Guilbert} A.,  {Alvarez-Candal} A.,  {Merlin} F.,  {Barucci} M.~A.,  {Dumas}
  C.,  {de Bergh} C.,   {Delsanti} A.,  2009, \mn@doi [\icarus]
  {10.1016/j.icarus.2008.12.023}, \href
  {http://adsabs.harvard.edu/abs/2009Icar..201..272G} {201, 272}

\bibitem[\protect\citeauthoryear{{Hall}, {Strobel}, {Feldman}, {McGrath}  \&
  {Weaver}}{{Hall} et~al.}{1995}]{Hall95}
{Hall} D.~T.,  {Strobel} D.~F.,  {Feldman} P.~D.,  {McGrath} M.~A.,   {Weaver}
  H.~A.,  1995, \mn@doi [\nat] {10.1038/373677a0}, \href
  {http://adsabs.harvard.edu/abs/1995Natur.373..677H} {373, 677}

\bibitem[\protect\citeauthoryear{{Hudson} \& {Moore}}{{Hudson} \&
  {Moore}}{2001}]{Hudson}
{Hudson} R.~L.,  {Moore} M.~H.,  2001, \mn@doi [\jgr] {10.1029/2000JE001299},
  \href {http://adsabs.harvard.edu/abs/2001JGR...10633275H} {106, 33275}

\bibitem[\protect\citeauthoryear{{Jewitt}}{{Jewitt}}{2009}]{Jewitt09}
{Jewitt} D.,  2009, \mn@doi [\aj] {10.1088/0004-6256/137/5/4296}, \href
  {http://adsabs.harvard.edu/abs/2009AJ....137.4296J} {137, 4296}

\bibitem[\protect\citeauthoryear{{Johnson}}{{Johnson}}{1990}]{Johnson1990}
{Johnson} R.~E.,  1990, {Energetic Charged-Particle Interactions with
  Atmospheres and Surfaces}

\bibitem[\protect\citeauthoryear{{Johnson}, {Lanzerotti}, {Brown},
  {Augustyniak}  \& {Mussil}}{{Johnson} et~al.}{1983}]{Johnson}
{Johnson} R.~E.,  {Lanzerotti} L.~J.,  {Brown} W.~L.,  {Augustyniak} W.~M.,
  {Mussil} C.,  1983, \aap, \href
  {http://adsabs.harvard.edu/abs/1983A%26A...123..343J} {123, 343}

\bibitem[\protect\citeauthoryear{{Johnson}, {Fam{\'a}}, {Liu}, {Baragiola},
  {Sittler}  \& {Smith}}{{Johnson} et~al.}{2008}]{Johnson2008}
{Johnson} R.~E.,  {Fam{\'a}} M.,  {Liu} M.,  {Baragiola} R.~A.,  {Sittler}
  E.~C.,   {Smith} H.~T.,  2008, \mn@doi [\planss] {10.1016/j.pss.2008.04.003},
  \href {http://adsabs.harvard.edu/abs/2008P%26SS...56.1238J} {56, 1238}

\bibitem[\protect\citeauthoryear{{Ka{\v n}uchov{\'a}}, {Brunetto}, {Melita}  \&
  {Strazzulla}}{{Ka{\v n}uchov{\'a}} et~al.}{2012}]{Kanuchova12}
{Ka{\v n}uchov{\'a}} Z.,  {Brunetto} R.,  {Melita} M.,   {Strazzulla} G.,
  2012, \mn@doi [\icarus] {10.1016/j.icarus.2012.06.043}, \href
  {http://adsabs.harvard.edu/abs/2012Icar..221...12K} {221, 12}

\bibitem[\protect\citeauthoryear{{Lunine}}{{Lunine}}{2006}]{Lunine}
{Lunine} J.~I.,  2006, \href
  {http://adsabs.harvard.edu/abs/2006mess.book..309L} {pp 309--319}

\bibitem[\protect\citeauthoryear{{Lv} et~al.,}{{Lv} et~al.}{2012}]{Lv}
{Lv} X.~Y.,  et~al., 2012, \mn@doi [\aap] {10.1051/0004-6361/201219886}, \href
  {http://adsabs.harvard.edu/abs/2012A%26A...546A..81L} {546, A81}

\bibitem[\protect\citeauthoryear{{Morbidelli}, {Levison}  \&
  {Gomes}}{{Morbidelli} et~al.}{2008}]{Morbidelli08}
{Morbidelli} A.,  {Levison} H.~F.,   {Gomes} R.,  2008, \href
  {http://adsabs.harvard.edu/abs/2008ssbn.book..275M} {pp 275--292}

\bibitem[\protect\citeauthoryear{{Muntean}, {Lacerda}, {Field}, {Fitzsimmons},
  {Hunniford}  \& {McCullough}}{{Muntean} et~al.}{2015}]{Muntean}
{Muntean} E.~A.,  {Lacerda} P.,  {Field} T.~A.,  {Fitzsimmons} A.,  {Hunniford}
  C.~A.,   {McCullough} R.~W.,  2015, \mn@doi [Surface Science]
  {10.1016/j.susc.2015.07.005}, \href
  {http://adsabs.harvard.edu/abs/2015SurSc.641..204M} {641, 204}

\bibitem[\protect\citeauthoryear{Petrik, Monckton, Koehler  \& Kimmel}{Petrik
  et~al.}{2014}]{petrik2014}
Petrik N.~G.,  Monckton R.~J.,  Koehler S. P.~K.,   Kimmel G.~A.,  2014,
  \mn@doi [The Journal of Chemical Physics]
  {http://dx.doi.org/10.1063/1.4878658}, 140, 204710

\bibitem[\protect\citeauthoryear{{Reisenfeld} et~al.,}{{Reisenfeld}
  et~al.}{2007}]{Reisenfeld}
{Reisenfeld} D.~B.,  et~al., 2007, \mn@doi [\ssr] {10.1007/s11214-007-9215-1},
  \href {http://adsabs.harvard.edu/abs/2007SSRv..130...79R} {130, 79}

\bibitem[\protect\citeauthoryear{{Schenk}, {Hamilton}, {Johnson}, {McKinnon},
  {Paranicas}, {Schmidt}  \& {Showalter}}{{Schenk} et~al.}{2011}]{Schenk2011}
{Schenk} P.,  {Hamilton} D.~P.,  {Johnson} R.~E.,  {McKinnon} W.~B.,
  {Paranicas} C.,  {Schmidt} J.,   {Showalter} M.~R.,  2011, \mn@doi [\icarus]
  {10.1016/j.icarus.2010.08.016}, \href
  {http://adsabs.harvard.edu/abs/2011Icar..211..740S} {211, 740}

\bibitem[\protect\citeauthoryear{{Shi}, {Baragiola}, {Grosjean}, {Johnson},
  {Jurac}  \& {Schou}}{{Shi} et~al.}{1995}]{Shi}
{Shi} M.,  {Baragiola} R.~A.,  {Grosjean} D.~E.,  {Johnson} R.~E.,  {Jurac} S.,
    {Schou} J.,  1995, \mn@doi [\jgr] {10.1029/95JE03099}, \href
  {http://adsabs.harvard.edu/abs/1995JGR...10026387S} {100, 26387}

\bibitem[\protect\citeauthoryear{{Sigmund}}{{Sigmund}}{1981}]{Sigmund}
{Sigmund} P.,  1981, {Sputtering by ion bombardment theoretical concepts}.
p.~9, \mn@doi{10.1007/3540105212_7}

\bibitem[\protect\citeauthoryear{{Strazzulla}, {Cooper}, {Christian}  \&
  {Johnson}}{{Strazzulla} et~al.}{2003a}]{Strazzulla03b}
{Strazzulla} G.,  {Cooper} J.~F.,  {Christian} E.~R.,   {Johnson} R.~E.,
  2003a, \mn@doi [Comptes Rendus Physique] {10.1016/j.crhy.2003.10.009}, \href
  {http://adsabs.harvard.edu/abs/2003CRPhy...4..791S} {4, 791}

\bibitem[\protect\citeauthoryear{{Strazzulla}, {Leto}, {Gomis}  \&
  {Satorre}}{{Strazzulla} et~al.}{2003b}]{Strazzulla03a}
{Strazzulla} G.,  {Leto} G.,  {Gomis} O.,   {Satorre} M.~A.,  2003b, \mn@doi
  [\icarus] {10.1016/S0019-1035(03)00100-3}, \href
  {http://adsabs.harvard.edu/abs/2003Icar..164..163S} {164, 163}

\bibitem[\protect\citeauthoryear{{Teolis}, {Vidal}, {Shi}  \&
  {Baragiola}}{{Teolis} et~al.}{2005}]{Teolis05}
{Teolis} B.~D.,  {Vidal} R.~A.,  {Shi} J.,   {Baragiola} R.~A.,  2005, \mn@doi
  [\prb] {10.1103/PhysRevB.72.245422}, \href
  {http://adsabs.harvard.edu/abs/2005PhRvB..72x5422T} {72, 245422}

\bibitem[\protect\citeauthoryear{Teolis, Shi  \& Baragiola}{Teolis
  et~al.}{2009}]{Teolis09}
Teolis B.,  Shi J.,   Baragiola R.,  2009, Journal of Chemical Physics, 130,
  134704

\bibitem[\protect\citeauthoryear{{Tiscareno} \& {Malhotra}}{{Tiscareno} \&
  {Malhotra}}{2003}]{Tiscareno03}
{Tiscareno} M.~S.,  {Malhotra} R.,  2003, \mn@doi [\aj] {10.1086/379554}, \href
  {http://adsabs.harvard.edu/abs/2003AJ....126.3122T} {126, 3122}

\bibitem[\protect\citeauthoryear{Vidal, Teolis  \& Baragiola}{Vidal
  et~al.}{2005}]{Vidal}
Vidal R.,  Teolis B.,   Baragiola R.,  2005, Surface Science, 588, 1

\bibitem[\protect\citeauthoryear{{Volk} \& {Malhotra}}{{Volk} \&
  {Malhotra}}{2013}]{Volk13}
{Volk} K.,  {Malhotra} R.,  2013, \mn@doi [\icarus]
  {10.1016/j.icarus.2013.02.016}, \href
  {http://adsabs.harvard.edu/abs/2013Icar..224...66V} {224, 66}

\bibitem[\protect\citeauthoryear{Westley, Baratta  \& Baragiola}{Westley
  et~al.}{1998}]{Westley}
Westley M.,  Baratta G.,   Baragiola R.,  1998, The Journal of chemical
  physics, 108, 3321

\bibitem[\protect\citeauthoryear{Wood \& Roux}{Wood \& Roux}{1982}]{Wood}
Wood B.,  Roux J.,  1982, JOSA, 72, 720

\bibitem[\protect\citeauthoryear{{von Steiger} et~al.,}{{von Steiger}
  et~al.}{2000}]{VonSteiger}
{von Steiger} R.,  et~al., 2000, \mn@doi [\jgr] {10.1029/1999JA000358}, \href
  {http://adsabs.harvard.edu/abs/2000JGR...10527217V} {105, 27217}

\makeatother
\end{thebibliography}

%%%%%%%%%%%%%%%%%%%%%%%%%%%%%%%%%%%%%%%%%%%%%%%%%%

%%%%%%%%%%%%%%%%%%%%%%%%%%%%%%%%%%%%%%%%%%%%%%%%%%

% Don't change these lines
\bsp	% typesetting comment
\label{lastpage}
\end{document}